\documentstyle[12pt]{article}
\begin{document}
\begin{center}
{\bf A stochastic model for the semiclassical collective dynamics of \\
     charged beams in particle accelerators}\footnote{To appear in
the Proceedings of the International Workshop on ``Quantum Aspects of
Beam Dynamics'', held in Stanford, 4--9 January 1998.} \\

\vspace{1.2cm}

Nicola Cufaro Petroni \\

\vspace{0.1cm}

{\it Dipartimento di Fisica, Universit\`a di Bari; 
                    and INFN, Sezione di Bari, \\
Via G. Amendola, Bari, Italia. E--Mail: cufaro@bari.infn.it} \\

\vspace{0.6cm}

Salvatore De Martino, Silvio De Siena and 
Fabrizio Illuminati$^{*}$ \\

\vspace{0.1cm}

{\it Dipartimento di Fisica, Universit\`a di Salerno; INFM, 
Unit\`a di Salerno; \\
and INFN, Sezione di Napoli -- Gruppo collegato di Salerno, I--84081 \\
Baronissi (Salerno), Italia. E--Mail: demartino@physics.unisa.it \\
desiena@physics.unisa.it; illuminati@physics.unisa.it}. \\
$*$ {\it Also at Fakult\"at f\"ur Physik, Universit\"at Konstanz, \\
Fach M 695, D--78457 Konstanz, Deutschland}

\end{center}

\abstract{In this paper we briefly review the main
aspects of a recent proposal to simulate semiclassical corrections
to classical dynamics by suitable classical stochastic fluctuations,
and we apply it to the specific instance of charged beams in particle
accelerators. The resulting picture is that the collective beam dynamics,
at the leading semiclassical order in Planck constant can be described
by a particular diffusion process, the Nelson process, that is
time--reversal invariant. Its diffusion coefficient $\sqrt{N}\lambda_c$ 
represents a semiclassical unit of emittance
(here $N$ is the number of particles in the beam, and $\lambda_c$ 
is the Compton wavelength). 
The stochastic dynamics of the Nelson
type can be easily recast in the form of a Schroedinger equation, with
the semiclassical unit of emittance replacing Planck constant. 
Therefore we provide a physical foundation to the several 
quantum--like models of beam dynamics proposed in recent years.
We also briefly touch upon applications of the Nelson and Schroedinger 
formalisms to incorporate the description of coherent collective effects.} 

\newpage
 
\section{Introduction}

The dynamical evolution of beams in particles accelerators is governed
by external electromagnetic forces and by the interaction of the beam
particles among themselves and with the environment. 
Charged beams are therefore higly nonlinear dynamical
systems, and most of the studies on colliding beams 
rely either on classical phenomena such as nonlinear resonances, 
or they are concerned with isolated sources of unstable behaviors
as building blocks of more complicated chaotic instabilities.

This line of inquiry has produced a general qualitative
picture of dynamical processes in particle accelerators
at the classical level.
However, the coherent oscillations of the beam density and 
profile require, to be explained, some mechanism of
local correlation and loss of statistical independence.
This fundamental observation points towards the need 
to take into account all the interactions as a whole. 
Moreover, the overall interactions
between charged particles and machine elements
are really nonclassical in the sense that of the
many sources of noise that are present, almost all
are mediated by fundamental quantum processes of
emission and absorbtion of photons.
Therefore the equations describing these processes 
must be, in principle, quantum.  

Starting from the above considerations,
two different approaches to the classical collective 
dynamics of charged beams have been developed, one
relying on the Fokker-Planck equation \cite{schonfeld}
for the beam density, another based on a mathematical
coarse graining of Vlasov equation leading to a
quantum--like Schroedinger equation, with a thermal
unit of emittance playing the role of Planck constant
\cite{fedele}.

The study of statistical effects on the dynamics
of electron (positron) colliding beams
by the Fokker--Planck equation has led to several interesting
results, and has become an established reference in treating
the sources of noise and dissipation in particle accelerators
by standard classical probabilistic techniques \cite{ruggiero}.

Concerning the relevance of the quantum--like approach, 
at this stage we only want to point out that some 
recent experiments on confined classical systems subject
to particular phase--space boundary conditions seem to
to be well explained by a quantum--like 
(Schroedinger equation) formalism \cite{varma}.

In any case, both approaches do not take into account quantum
corrections, while in principle these effects 
should be relevant, expecially in fixing fundamental lower 
limits to beam emittance.
In this report we give a short summary of
a recently proposed model for the description
of collective beam dynamics in the semiclassical regime. 
This new approach relies on the idea of
simulating semiclassical corrections to
classical dynamics by suitable classical stochastic 
fluctuations with long range coherent correlations,
whose scale is ruled by Planck constant \cite{demartino}.

The fluctuative hypothesis has been introduced 
by simple stability criteria, and it has been semiquantitatively 
tested for many stable systems, including beams.
The virtue of the proposed semiclassical model is twofold:
on the one hand it can be formulated both in a probabilistic 
(Fokker--Planck) fashion and in a quantum--like 
(Schroedinger) setting. It thus bridges the formal
gap between the two approaches. 
At the same time it goes further by
describing collective effects beyond the classical regime due 
to the semiclassical quantum corrections. 

In particular, implementing
the fluctuative hypothesis qualitatively
by simple dimensional analysis, we derive
a formula for the phase--space unit of emittance that 
connects in a nontrivial way the number of particles in the 
beam with Plank constant.

The fluctuative scheme is then implemented
quantitatively by introducing a random kinematics
in the form of a diffusion process in configuration
space for a generic representative of the
beam (collective degree of freedom).

We are interested in the description of the stability
regime, when thermal dissipative effects are
balanced on average by the RF energy pumping, and the
overall dynamics is conservative and time--reversal
invariant in the mean. Therefore, we model the random
kinematics with a particular class of diffusion processes,
the Nelson diffusions, that are nondissipative and 
time--reversal invariant (We will briefly comment at the
end of the last section on the extension of the present scheme 
to include the treatment of dissipative effects).

The diffusion process describes the effective motion
at the mesoscopic level (interplay of thermal equilibrium,
classical mechanical stability, and fundamental quantum 
noise) and therefore the diffusion coefficient is set to be 
the semiclassical unit of emittance provided by qualitative 
dimensional analysis. In other words, we simulate the quantum
corrections to classical deterministic motion (at leading order
in Planck constant) with a suitably defined random kinematics
replacing the classical deterministic trajectories.

Finally, the dynamical equations are derived via the variational
principle of classical dynamics, with the only crucial
difference that the kinematical rules and the dynamic
quantities, such as the Action and the Lagrangian, are
now random. 
The stochastic variational principle leads to a pair
of coupled equations for the beam density and the beam
center current velocity, describing the dynamics of
beam density oscillations. It is an effective description
in the stability regime.

The stochastic variational principle for Nelson
diffusions (with diffusion coefficient equal to Planck constant)
is a well developed mathematical tool that has
originally been introduced to provide a stochastic formulation of
quantum mechanics. Therefore, apart from the different objects
involved (beam spatial density versus Born probability density;
Planck constant versus emittance), the dynamical equations of 
our model formally reproduce the equations of the Madelung
fluid (hydrodynamic) representation of quantum mechanics. 
In this sense, the present scheme allows for a 
quantum--like formulation equivalent to the probabilistic one.

At the end of the last section we will briefly discuss how the 
hydrodynamic formulation of the equations for the collective 
stochastic dynamics can be used to control the beams, 
for instance by selecting the form of the external potential 
needed to obtain coherent oscillations of the beam density.
 
\section{Simulation of semiclassical effects 
by classical fluctuations}

Let us consider a physical system subject to a classical
force law of modulus $F(r)$ that is attractive and confining at
least for some finite space region with a characteristic
linear dimension $R$. Given $N$ elementary granular constituents
of the system, each of mass $m$, let $v$ denote their 
characteristic velocity, and $\tau$ their characteristic unit of
time.

A characteristic unit of action per particle is then defined as
\begin{equation}
\alpha=mv^{2}\tau.
\end{equation}

If the system has to be stable and confined, one must impose
that the characteristic potential energy of each particle
be on average equal to its characteristic kinetic energy
(virial theorem):
\begin{equation}      
{\cal{L}} \cong mv^{2},
\end{equation}

\noindent where $\cal{L}$ is the work performed by the
system on a single constituent. On the other hand, if the system
extends on the characteristic length scale $R$, 
\begin{equation}
{\cal{L}} \cong NF(R)R.
\end{equation}

\noindent By equations (2) and (3) we can express the 
characteristic velocity $v$ as
\begin{equation}
v \cong \sqrt{ \frac{NF(R)R}{m}}.
\end{equation}

\noindent Introducing the global time 
scale ${\cal{T}}$ associated to the system,
we also have $v=R/{\cal{T}}$. Replacing this expression
and equation (4) for each power of $v$ in equation (1),
we obtain the following expression for the
action per particle:
\begin{equation}
\alpha \cong \sqrt{mF(R)} R^{3/2} N^{1/2}
\frac{\tau }{\cal{T}}.
\end{equation}

\noindent Mechanical stability requires that the action per 
particle must not depend on $N$, while on the other hand, the 
microscopic unit of time $\tau$ must obviously depend on 
$N$ and on the system's global time scale ${\cal{T}}$.
Therefore we must impose 
\begin{equation}
\tau = \frac{{\cal{T}}}{\sqrt{N}}.
\end{equation}

\noindent Inserting equation (6) into equation (5) we obtain the
unit of action per particle as a explicit expression in
terms of the constituent's mass, the system's linear dimension
$R$, and the classical force calculated in $R$: 
\begin{equation}
\alpha\cong m^{1/2}R^{3/2} \sqrt{F(R)}.
\end{equation}

The scaling relation (6) can be also interpreted 
as a fluctuative hypothesis connecting the time scale
of a microscopic stochastic motion with the classical
time scale of the global system. In fact, equation (6)
was first postulated by F. Calogero 
in his attempt to prove that quantum mechanics might
be interpreted as a tiny chaotic component of 
the individual particles' motion in a gravitationally
interacting Universe \cite{calogero}.

In our scheme, rather than being a postulated consequence
of classical gravitational chaos, the fluctuative
hipothesis of Calogero derives from a condition of
mechanical stability. Since the stability conditions
and the virial theorem apply to any classically stable
and confined system, even with a small number of degrees of
freedom, our derivation of equations (6) and (7) is universal
as it applies to any interactions, not only gravity, and to
systems composed by any number of constituents, not necessarily
large, and not necessarily classically chaotic.

We have verified that for any stable aggregate, plugging in
equation (7) the pertaining interaction $F$, 
individual constituents' mass $m$
and aggregate's linear dimension $R$, one has that
{\it the unit of action per particle} $\alpha$ {\it is always
equal, in order of magnitude, to Planck action constant} $h$. 

Our interpretation of this remarkable result is then that the 
fluctuative relation (6) and the associated formula (7) for 
the Planck quantum of action
simulate (reformulate) in a classical probabilistic
language the Bohr-Sommerfeld quantization condition. They
provide a classical description of quantum corrections to
classical phase--space dynamics at the leading semiclassical
order $h$.

We here briefly derive the result for the case of interest
of a stable bunch of charged particles in a particle accelerator.
We consider a single electron (proton), in the
reference frame comoving with the bunch.
Confinement and stability of the bunch arise from the many
complicated interactions among its constituents and between
the same constituents and the external magnetic and RF fields.
The net effect can be, in first approximation, schematized by
saying that the single electron (proton)
experiences an effective harmonic
force, the typical phenomenological 
law of force for beams when higher anharmonic contributions 
can be neglected: $F(R)\cong KR$, where $K$ is the effective 
phenomenological elastic constant. We then have for beams:
\begin{equation}
\alpha\cong m^{1/2} R^{2} K^{1/2}.
\end{equation}

Let us consider for instance the transverse oscillations
for protons at Hera: in this case we have $K=10^{-12}Nm^{-1}$, 
the linear transverse dimension of the bunch $R=10^{-7}m$, 
and the proton mass.
For electrons in linear colliders we have instead
$K=10^{-11}Nm^{-1}$, $R=10^{-7}m$, and the electron mass.
In both cases, from equation (8) we have that the unit of
action per particle $\alpha$, ruling the coherence and 
stability of the bunch, is in both cases $h$, up to at
most one order of magnitude. 

All other instances of charged bunches considered
lead to the same result, yielding our first important
conclusion: the stability of charged beams is ruled by 
quantum effects on a mesoscopic scale. Moreover, at the 
semiclassical level, such quantum aspects can be described
in terms of suitable classical fluctuations that mimick
(simulate) the weak but unavoidable presence of fundamental
quantum noise.
  
The parameter that rules the stability of
the system at the mesoscopic scale is however not directly
$h$, but in the case of charged beams some characteristic
unit of emittance. This is a scale of action, or of
length when divided by the Compton wavelength, 
that measures the spread of the bunch in phase space, 
or, equivalently, in real space.

This notion is very useful in the regime of stability and of
thermal equilibrium that we explicitely consider. In this
case the emittance can be expressed as
a unit of equivalent thermal action.
To introduce a characteristic unit of emittance
in our fluctuative semiclassical scheme we then proceed
as follows: the time scale of quantum fluctuations is 
defined as the ratio between $h$ and a suitable energy
describing the equilibrium state of the
given system. 
This leads naturally to identify this energy with
the equivalent thermal energy $k_{B}T$, with $k_{B}$ the
Boltzmann constant and $T$ the equivalent temperature.
On the other hand, in our scheme such time scale
coincides with the fluctuative time $\tau$; we
therefore have:
\begin{equation}
\tau \cong \frac{h}{k_{B}T} \, .
\end{equation}

\noindent Using relation (6) we obtain the equivalent thermal
unit of action  
\begin{equation}
k_{B}T{\cal{T}} \cong h\sqrt{N} \, .
\end{equation}

\noindent Introducing the Compton wavelength 
$\lambda_{c} = h/mc$ and dividing by it both 
sides of equation (10) we finally obtain the
characteristic unit of emittance ${\cal{E}}$:
\begin{equation}
{\cal{E}} \cong \lambda_{c}\sqrt{N} \, .
\end{equation}

Equation (11) connects in a nontrivial way the
number of particles in a given charged beam and
the Compton wavelength. The
square root of $N$ appears as 
a semiclassical ``memory'' of quantum 
interference. The relation (11) seems to
point out the existence of a mesoscopic
lower bound on the emittance some orders of
magnitude above the quantum limit 
given by the Compton wavelength. Moreover,
Equation (11) yields the correct order of 
magnitude in for the emittance in typical
accelerators: for instance, with $N \cong 10^{11}
\div 10^{-12}$, one has ${\cal{E}} \cong 10^{-6}m$ 
in excellent agreement with the lowest emittances
that are at the moment experimentally attainable.  

Actually, limits and requirements
on beam existence, luminosity and statistics 
do not allow for beams with a number of particles 
appreciably lower than $N \cong 10^{10} \div 10^{11}$.
Thus the estimate (11) really provides an 
{\it a priori} lower bound, as it implies 
that the emittance cannot be reduced appreciably
below the mesoscopic thresholds ${\cal{E}} \cong
10^{5} \div 10^{6} \lambda_{c}$, well above
the Compton wavelength limit and only one or two  
orders of magnitude below the current experimental
limits. It seems also unlikely that
further quantum corrections beyond the leading
semiclassical order could somehow contribute in
lowering the mesoscopic bound (11) as a function of $N$.

\section{Stochastic collective dynamics in the stability regime}

The previous discussion can be made more quantitative
by observing that the fluctuative relation (6) can be
be recast with a little bit of work in the alternative form
\begin{equation} 
l \sim \tau^{2/3} \, ,
\end{equation}

\noindent where $l$ is a characteristic mean free path
per particle. The detailed derivation of relation (12) from 
equation (6) is reported elsewhere \cite{demartino}.
Relation (12) indicates that the classical
flcutuative simulation of semiclassical corrections
really implies a fractal space--time relation in the
mean, with a Kepler exponent associated to stable,
confined and coherent dynamical systems, for instance
charged beams in the stability regime.

We therefore model the spatially coherent
fluctuations (6) and (12) by a random kinematics
performed by some collective degree of freedom
$q(t)$ representative of the beam. The most universal
continuous random kinematics that we can choose is  
a diffusion process in real or configuration space.
In this way the random kinematics provides an effective
description of the space--time variations of the particle
beam density $\rho(x,t)$ as it coincides with the
probability density of the diffusion process performed
by $q(t)$.

Since it measures a collective effect at the mesoscopic scale,
the diffusion coefficient must be related to the equilibrium
parameter in the stability regime, that is to the characteristic
semiclassical unit of emittance (11) rather than to the
Plank action constant.

Then, in suitable units, the basic stochastic kinematical
relation is the Ito stochastic differential equation
\begin{equation}
 dq(t) = v_{+}(q,t)dt + {\cal{E}}^{1/2}dw \, ,
\end{equation}

\noindent where $v_{+}$ is the deterministic drift, the square
root of the characteristic emittance (11) is the diffusion 
coefficient, and $dw$ is the time increment of the standard
$\delta$--correlated Wiener noise.

We are concerned with the regime of stability of the beam
oscillation dynamics, both since it is the relevant regime 
in the physics of accelerators and because the beam can be 
considered quasistationary during it, until, eventually, 
space charge effects become dominant and the beam is lost.
In such stationary regime the energy lost by
photonic emissions is regained in the RF cavities, and
on average the dynamics is still time--reversal invariant.
We can therefore still define a classical Lagrangian 
$L(q, {\dot{q}})$ for the system, however with the 
classical deterministic kinematics replaced by the random
diffusive kinematics (13). 

The equations of dynamics can then be deduced from the classical
Lagrangian by simply modifying the variational principles of
classical mechanics into stochastic variational principles. 
In fact, the mathematical techniques of stochastic variational
principles have been developed and applied to obtain Nelson 
stochastic mechanics, an independet stochastic reformulation 
of quantum mechanics in terms of time--reversal 
invariant Markov diffusion processes with diffusion
coefficient $\hbar$ \cite{nelson}. In the context of 
Nelson stochastic mechanics one derives Schroedinger
equation in the form of the Madelung coupled hydrodynamic
equations for the probability density and the probability 
current \cite{nelson}.
 
In the present mesoscopic context the analysis is quite 
similar to that of Nelson stochastic mechanics, yielding
again two coupled nonlinear hydrodynamic equations,
however, with the emittance (11) replacing Planck constant 
in the diffusion coefficient, the real space bunch density 
replacing the quantum mechanical probability density, 
and the bunch center velocity replacing the quantum mechanical 
probability current. 

We now briefly sketch the derivation of the dynamical equations.
The detailed analysis may be found elsewhere \cite{futuro}.
Given the stochastic differential equation (13) for the diffusion
process $q(t)$ in $d=3$ space dimensions, one introduces
the classical Lagrangian
\begin{equation}
L(q, {\dot q})=1/2m {\dot q^{2}}- V(q) \, .
\end{equation}

For the generic trial diffusion $q(t)$ one has, respectively,
the probability density $\rho(x,t)$, the forward drift $v_{+}(x,t)$ 
and the backward drift $v_{-}(x,t)$. It is then useful to define
two new variables, $v(x,t)$ and $u(x,t)$, respectively the current
velocity and the osmotic velocity, defined as:
\begin{equation}
v=\frac{v_{+} + v_{-}}{2} \; ; \; \; \; 
u=\frac{v_{+}-v_{-}}{2} = {\cal{E}}\frac{\nabla \rho}{\rho} \, . 
\end{equation}

The mean classical action is defined in strict analogy to the
classical action in the deterministic case, but for the limiting
procedure that needs to be taken in the sense of expectation
values, as the sample paths of a diffusion process are non
differentiable:
\begin{equation}
A(t_{0}, t_{1}; q) = \int_{t_{0}}^{t_{1}}
lim_{\Delta t \rightarrow 0^{+}}
E \left[ \frac{m}{2} \left( \frac{\Delta q}{\Delta t}
\right)^{2} - V(q) \right] dt \, ,
\end{equation}

\noindent where $E$ denotes the expectation with respect to the
probability density $\rho$. It can be shown that the mean classical
action (16) associated to the diffusive kinematics (13) can be
cast in the following particularly appealing Eulerian
hydrodynamic form \cite{nelson}:
\begin{equation}
A(t_{0}, t_{1}; q) = \int_{t_{0}}^{t_{1}}dt\int d^{3}x
\left[ \frac{m}{2} \left( v^{2} -u^{2} \right) - V(x) \right]
\rho(x,t) \, .
\end{equation}

The stochastic variational principle now follows: the Action
is stationary, $\delta A = 0$, under smooth variations of
the density $\delta \rho$, and of the current velocity
$\delta v$, with vanishing boundary conditions at the initial
and final times, if and only if the current velocity is the
gradient of some scalar field $S(x,t)$ (the phase):
\begin{equation}
mv = \nabla S \; .
\end{equation}

With the above conditions met, the two coupled nonlinear Lagrangian
equations of motion for the density $\rho$ (or alternatively for
the osmotic velocity $u$) and for the current velocity $v$ (or
alternatively for the phase $S$) are the Hamilton--Jacobi--Madelung
equation:
\begin{equation}
\partial_{t}S + \frac{m}{2}v^{2} - 2m{\cal{E}}^{2}
\frac{\nabla^{2}\sqrt\rho}{\sqrt\rho} + V(x) = 0 \, ,
\end{equation}   

\noindent and the continuity equation:
\begin{equation}
\partial_{t}\rho = -\nabla [\rho v].
\end{equation} 

By solving equations (19) and (20) the state of the bunch is
completely determined. Formal linearization of the 
equations can be achieved through the standard De Broglie
ansatz yielding the Schroedinger equation of 
the quantum--like models. However, one should bear in mind
that the real hydrodynamic equations (19)--(20) are the 
physically fundamental objects, while linearizing them to a complex
Schroedinger equation is a bare mathematical tool that can 
be useful for calculational needs, but bears no physical 
significance. In particular, the complex wave function is
devoid of any physical meaning. Thus, in the present context, 
the situation is just the opposite to that in quantum mechanics, 
where instead the wave function and the Schroedinger equation are
the fundamental physical ingredients.

The observable structure is quite clear:
the expectations (first moments) of the three components of the 
current velocity $v$ are the average velocities of oscillation of 
the bunch center along the longitudinal and transverse directions.
The expectations (first moments) of the three components of the
process $q(t)$ give the average coordinate of the bunch center.
The second moments of $q(t)$ allow to determine the 
dispersion (spreading) of the bunch. In the harmonic case, these
are all the moments that are needed (Gaussian probability density),
and we have coherent state solutions. In the anharmonic case the
coupled equations of dynamics may be used to achieve a controlled
coherence: given a desired state $(\rho, v)$ the equations of motion
(19) and (20) can be solved for the external controlling potential 
$V(x,t)$ that realizes the desired state. Lack of space prevents us
from commenting further on this very important application of our
formalism. A thorough and detailed study of the controlled coherent 
evolutions in the framework of our stochastic model will
be presented in a forthcoming paper \cite{futuro2}.

\section{Aknowledgement}

One of us (F.I.) gratefully aknowledges a research
fellowship from the Alexander von Humboldt Stiftung and hospitality 
from the LS Mlynek at the Fak\"ult\"at f\"ur Physik of the University
of Konstanz.

\end{document}